\documentclass[a4paper]{article}
\usepackage[utf8]{inputenc}
\usepackage[T1]{fontenc}
\usepackage{amsmath,amssymb,array}
\usepackage{booktabs}

\usepackage{graphicx}

\providecommand{\tightlist}{%
  \setlength{\itemsep}{0pt}\setlength{\parskip}{0pt}}

\RequirePackage{geometry}
\RequirePackage{fancyhdr}
\RequirePackage{fancyvrb}
\RequirePackage{alltt}

\RequirePackage[hyphens]{url}
\RequirePackage[pagebackref]{hyperref}

\let\pkg=\strong
\newcommand{\CRANpkg}[1]{\href{http://CRAN.R-project.org/package=#1}{\pkg{#1}}}%

\usepackage{natbib}

\newcommand{\address}[1]{\addvspace{\baselineskip}\noindent\emph{#1}}

\makeatletter
\renewcommand{\@seccntformat}[1]{}
\makeatother

\RequirePackage{xcolor}
\definecolor{link}{rgb}{0.45,0.51,0.67}
\hypersetup{
  colorlinks,%
  citecolor=link,%
  filecolor=link,%
  linkcolor=link,%
  urlcolor=link
}

\DefineVerbatimEnvironment{Sinput}{Verbatim}{fontsize=\small}
\DefineVerbatimEnvironment{Soutput}{Verbatim}{fontsize=\small}
\DefineVerbatimEnvironment{Scode}{Verbatim}{fontsize=\small}
\DefineVerbatimEnvironment{Sin}{Verbatim}{fontsize=\small}
\DefineVerbatimEnvironment{Sout}{Verbatim}{fontsize=\small}
\newenvironment{Schunk}{}{}

\begin{document}

\title{Computer Algebra in R with \CRANpkg{caracas}}
\author{by Mikkel Meyer Andersen and Søren Højsgaard}

\maketitle

\abstract{%
The capability of R to do symbolic mathematics is enhanced by the
\CRANpkg{caracas} package. This package uses the Python computer algebra
library SymPy as a back-end but \CRANpkg{caracas} is tightly integrated
in the R environment, thereby enabling the R user with symbolic
mathematics within R. Key components of the \CRANpkg{caracas} package
are illustrated in this paper. Examples are taken from statistics and
mathematics. The \CRANpkg{caracas} package integrates well with
e.g.~\CRANpkg{Rmarkdown}, and as such creation of scientific reports and
teaching is supported.
}

\newlength{\fancyvrbtopsep}
\newlength{\fancyvrbpartopsep}
\makeatletter
\FV@AddToHook{\FV@ListParameterHook}{\topsep=\fancyvrbtopsep\partopsep=\fancyvrbpartopsep}
\makeatother

\setlength{\fancyvrbtopsep}{0pt}

\RecustomVerbatimEnvironment{Sinput}{Verbatim}{xleftmargin=3mm,formatcom=\color{black}}

\RecustomVerbatimEnvironment{Soutput}{Verbatim}{xleftmargin=4mm,formatcom=\color{black}}

\def\EE{\mathbf{E}}
\def\var{\mathbf{Var}}
\def\tr{\mathbf{tr}}
\def\det{\mathbf{det}}
\def\diag{\mathbf{diag}}
\def\sympy{"SymPy"}

\def\inv{^{-1}}
\def\transp{^\top}
\def\cip{\perp\!\!\perp}

\newcommand{\matrxr}[1]
{\left(
    PCA\begin{array}{rrrrrrrrrrrrrrrrrrrrrrrrrrrrrrrrrrrrr}
      #1 \\
    \end{array}
  \right)}

\newcommand{\matrxc}[1]
{\left(
    \begin{array}{cccccccccccccccccccccccccccccccccccc}
      #1 \\
    \end{array}
  \right)}

\parindent0pt

\hypertarget{keywords}{%
\subsection{Keywords}\label{keywords}}

Differentiation, Factor analysis, Hessian matrix, Integration, Lagrange
multiplier, Limit, Linear algebra, Principal component analysis, Score
function, Symbolic mathematics, Taylor expansion, Teaching.

\hypertarget{introduction}{%
\subsection{Introduction}\label{introduction}}

The capability of R \citep{R} to handle symbolic mathematics is greatly
enhanced by two packages: The \CRANpkg{caracas} package, which is the
main topic of this paper, and the \CRANpkg{Ryacas} package described in
\citet{ryacas}. The \CRANpkg{caracas} package is based on interfacing
the Python library SymPy \citep{sympy}, using the \CRANpkg{reticulate}
package, \citep{reticulate}. Similarly, \CRANpkg{Ryacas} is based on
interfacing the computer algebra system (CAS) yacas
\citep{yacas, Pinkus2002}. The \CRANpkg{caracas} package is open-source
and the source code is available at
\url{https://github.com/r-cas/caracas}. Several vignettes illustrating
\CRANpkg{caracas} are provided and these are also available online, see
\url{https://r-cas.github.io/caracas/}.

One particular instance where we have found the packages useful is in
connection with teaching where symbolic mathematics is helpful strongly
aided by the packages ability to enter in a reproducible framework
(provided e.g.~by \CRANpkg{Rmarkdown}). In this paper we provide a few
examples of this, and we address the issue more generally towards the
end of the paper.

\hypertarget{the-package-versus-other-computer-algebra-systems}{%
\paragraph{\texorpdfstring{The \CRANpkg{caracas} package versus other
computer algebra
systems}{The  package versus other computer algebra systems}}\label{the-package-versus-other-computer-algebra-systems}}

Neither \CRANpkg{caracas} nor \CRANpkg{Ryacas} are as powerful as some
of the large commercial computer algebra systems. The virtue of the
\CRANpkg{caracas} and \CRANpkg{Ryacas} packages lie elsewhere:

\begin{enumerate}
\def\labelenumi{\arabic{enumi}.}
\item
  Tools like solving equations, summation, limits, symbolic linear
  algebra, outputting in tex format etc. are directly available from
  within R.
\item
  The packages enable working with the same language and in the same
  environment as the user does for statistical analyses.
\item
  Symbolic mathematics can easily be combined with data which is helpful
  in e.g.~numerical optimization.
\item
  Lastly, the packages are part of the R project (since the packages are
  on CRAN). As such the packages are freely available, and therefore
  support e.g.~education - also of people with limited economical means
  and thus contributing to United Nations sustainable development goals,
  cfr. \citet{UN17}.
\end{enumerate}

With respect to freely available resources in a CAS context, we would
like to draw attention to \texttt{WolframAlpha}, see
\url{https://www.wolframalpha.com/}, which is an online for answering
(mathematical) queries.

\hypertarget{introductory-examples}{%
\subsection{Introductory examples}\label{introductory-examples}}

There are no other system requirements than Python for using
\CRANpkg{caracas}. This paper is based on the following version of
\CRANpkg{caracas}:

\begin{Schunk}
\begin{Sinput}
R> library(caracas)
R> packageVersion("caracas")
\end{Sinput}
\begin{Soutput}
#> [1] '1.1.0'
\end{Soutput}
\end{Schunk}

\hypertarget{the-interplay-between-r-and-sympy}{%
\subsubsection{The interplay between R and
SymPy}\label{the-interplay-between-r-and-sympy}}

As mentioned above, \CRANpkg{caracas} provides an interface from R to
the Python package SymPy. This means that SymPy is ``running under the
hood'' of R via the \CRANpkg{reticulate} package. In \CRANpkg{caracas}
we have symbols, which is an R list with a \texttt{pyobj} slot and the
class \texttt{caracas\_symbol}. The \texttt{pyobj} refers to an object
in Python (often a SymPy object). As such, a symbol (in R) provides a
handle to a Python object. In the design of \CRANpkg{caracas} we have
tried to make this distinction something the user should not be
concerned with, but it is worthwhile being aware of the distinction.
There are several ways of creating symbols; one is with
\texttt{def\_sym()} that both declares the symbol in R and in Python:

\begin{Schunk}
\begin{Sinput}
R> ## Define symbols and assign in global environment
R> def_sym(s1, s2); s1 # Declares 's1'/'s2' in both R and Python
\end{Sinput}
\begin{Soutput}
#> [caracas]: s1
\end{Soutput}
\begin{Sinput}
R> str(s1)
\end{Sinput}
\begin{Soutput}
#> List of 1
#>  $ pyobj:s1
#>  - attr(*, "class")= chr "caracas_symbol"
\end{Soutput}
\begin{Sinput}
R> ## Create new symbol from existing ones
R> s3 <- s1 * s2; s3 # 's3' is a symbol in R; no corresponding object in Python
\end{Sinput}
\begin{Soutput}
#> [caracas]: s1*s2
\end{Soutput}
\begin{Sinput}
R> str(s3)
\end{Sinput}
\begin{Soutput}
#> List of 1
#>  $ pyobj:s1*s2
#>  - attr(*, "class")= chr "caracas_symbol"
\end{Soutput}
\end{Schunk}

Note that above \texttt{def\_sym(s1,\ s2)} is a short-hand for the
following:

\begin{Schunk}
\begin{Sinput}
R> s1 <- symbol("s1")
R> s2 <- symbol("s2")
\end{Sinput}
\end{Schunk}

We can further exemplify that objects in R and Python are not
necessarily identical. We look into a symbol, and to make the
distinction clear we use different names. Symbols can be created with
the \texttt{symbol()} function.

\begin{Schunk}
\begin{Sinput}
R> ## Create a symbol 'b1' corresponding to an entity called 'a' in SymPy:
R> b1 <- symbol("a"); str(b1)
\end{Sinput}
\begin{Soutput}
#> List of 1
#>  $ pyobj:a
#>  - attr(*, "class")= chr "caracas_symbol"
\end{Soutput}
\begin{Sinput}
R> ## A new symbol can be created as:
R> b2 <- b1 + 1; str(b2)
\end{Sinput}
\begin{Soutput}
#> List of 1
#>  $ pyobj:a + 1
#>  - attr(*, "class")= chr "caracas_symbol"
\end{Soutput}
\begin{Sinput}
R> ## The Python entity 'a' in the symbol can be modified with:
R> b3 <- subs(b2, "a", "k"); str(b3)
\end{Sinput}
\begin{Soutput}
#> List of 1
#>  $ pyobj:k + 1
#>  - attr(*, "class")= chr "caracas_symbol"
\end{Soutput}
\end{Schunk}

Going back to the first example, we can substitute one symbol with
another and simplify the result as (where we use the pipe operator
\texttt{\%\textgreater{}\%} from \CRANpkg{magrittr} by
\citet{magrittr}):

\begin{Schunk}
\begin{Sinput}
R> s4 <- s3 %>% subs("s1", "u + v") %>% subs("s2", "u - v"); s4
\end{Sinput}
\begin{Soutput}
#> [caracas]: (u - v)*(u + v)
\end{Soutput}
\begin{Sinput}
R> s5 <- expand(s4); s5
\end{Sinput}
\begin{Soutput}
#> [caracas]:  2    2
#>            u  - v
\end{Soutput}
\end{Schunk}

It is also possible to convert to and from symbols and standard R
expressions:

\begin{Schunk}
\begin{Sinput}
R> ## Coerce from symbol to expression:
R> e5 <- as_expr(s5); e5
\end{Sinput}
\begin{Soutput}
#> expression(u^2 - v^2)
\end{Soutput}
\begin{Sinput}
R> ## Coerce from expression to symbol:
R> as_sym(e5) # identical to s5
\end{Sinput}
\begin{Soutput}
#> [caracas]:  2    2
#>            u  - v
\end{Soutput}
\end{Schunk}

\hypertarget{finding-a-limit---the-euler-constant}{%
\subsubsection{Finding a limit - the Euler
constant}\label{finding-a-limit---the-euler-constant}}

Define symbols \texttt{n} and \texttt{f}:

\begin{Schunk}
\begin{Sinput}
R> def_sym(n)
R> f <- (1 + 1/n)^n
\end{Sinput}
\end{Schunk}

We can calculate the limit of \texttt{f} for \(n \to \infty\):

\begin{Schunk}
\begin{Sinput}
R> lim_f <- lim(f, n, Inf)
R> lim_f
\end{Sinput}
\begin{Soutput}
#> [caracas]: exp(1)
\end{Soutput}
\begin{Sinput}
R> as_expr(lim_f)
\end{Sinput}
\begin{Soutput}
#> [1] 2.72
\end{Soutput}
\end{Schunk}

We can also tell \CRANpkg{caracas} not to evaluate the limit (with the
\texttt{doit\ =\ FALSE} argument) but only set up the symbol for later
evaluation and/or for additional algebraic manipulations:

\begin{Schunk}
\begin{Sinput}
R> lim_f_sym <- lim(f, n, Inf, doit = FALSE)
R> lim_f_sym
\end{Sinput}
\begin{Soutput}
#> [caracas]:             n
#>                 /    1\ 
#>             lim |1 + -| 
#>            n->oo\    n/
\end{Soutput}
\end{Schunk}

By default \CRANpkg{caracas} uses UTF-8 printing, but in this paper we
have used pretty ASCII printing which can be set globally by
\texttt{options(caracas.print.prettyascii\ =\ TRUE)}.

The unevaluated symbol can be evaluated as follows:

\begin{Schunk}
\begin{Sinput}
R> lim_f <- doit(lim_f_sym) 
R> lim_f
\end{Sinput}
\begin{Soutput}
#> [caracas]: exp(1)
\end{Soutput}
\end{Schunk}

Hence, three \CRANpkg{caracas} symbols have been created above:
\texttt{f}, \texttt{lim\_f\_sym} and \texttt{lim\_f}. Objects can be
printed in \TeX~form using \texttt{tex()}, e.g.

\begin{Schunk}
\begin{Sinput}
R> tex(lim_f_sym)
\end{Sinput}
\begin{Soutput}
#> [1] "\\lim_{n \\to \\infty} \\left(1 + \\frac{1}{n}\\right)^{n}"
\end{Soutput}
\end{Schunk}

This can be used in a \TeX~environment as e.g.

\begin{Schunk}
\begin{Sinput}
\[
`r tex(f)`, \quad `r tex(lim_f_sym)`, \quad `r tex(lim_f)`.
\]
\end{Sinput}
\end{Schunk}

giving \[
\left(1 + \frac{1}{n}\right)^{n}, \quad \lim_{n \to \infty} \left(1 + \frac{1}{n}\right)^{n}, \quad e.
\]

\hypertarget{differentiation-and-integration}{%
\subsubsection{Differentiation and
integration}\label{differentiation-and-integration}}

Consider this function (taken from a vignette for the
\CRANpkg{mosaicCalc} package \citep{mosaicCalc}). Using the
\CRANpkg{Deriv} package, the derivative can be found as follows:

\begin{Schunk}
\begin{Sinput}
R> f <- function(x){
+   a * x + b * x^2 + c * sin(x^2)
+ }
R> Deriv::Deriv(f, "x")
\end{Sinput}
\begin{Soutput}
#> function (x) 
#> a + x * (2 * (c * cos(x^2)) + 2 * b)
\end{Soutput}
\end{Schunk}

The anti-derivative, however, is not easily obtained in R. Using
\CRANpkg{caracas} we get derivative and anti-derivative as:

\begin{Schunk}
\begin{Sinput}
R> f_c <- as_sym("a * x + b * x^2 + c * sin(x^2)")
R> def_sym(x) # To get handle on x in R
R> D_f <- der(f_c, x) # Or: der(f_c, "x")
R> aD_f <- int(f_c, x) %>% simplify()
\end{Sinput}
\end{Schunk}

\[
D_f = a + 2 b x + 2 c x \cos{\left(x^{2} \right)}, \quad aD_f = \frac{a x^{2}}{2} + \frac{b x^{3}}{3} + \frac{\sqrt{2} \sqrt{\pi} c S\left(\frac{\sqrt{2} x}{\sqrt{\pi}}\right)}{2}
\]

Above, \(S()\) is the Fresnel integral
\(S(z) = \int_0^z \sin \left (\frac{\pi}{2}t^2 \right ) dt\). Evaluation
in R requires a definition of \texttt{fresnels()}:

\begin{Schunk}
\begin{Sinput}
R> as_expr(aD_f) # Evaluation requires user-defined fresnels()
\end{Sinput}
\begin{Soutput}
#> expression(a * x^2/2 + b * x^3/3 + sqrt(2) * sqrt(pi) * c * fresnels(sqrt(2) * 
#>     x/sqrt(pi))/2)
\end{Soutput}
\end{Schunk}

\hypertarget{exact-and-numerical-evaluations}{%
\subsubsection{Exact and numerical
evaluations}\label{exact-and-numerical-evaluations}}

We can make exact as well as numerical evaluations as follows:

\begin{Schunk}
\begin{Sinput}
R> def_sym(x)
R> f <- exp(x^2)
R> subs(f, x, "1/3")
\end{Sinput}
\begin{Soutput}
#> [caracas]: exp(1/9)
\end{Soutput}
\begin{Sinput}
R> subs(f, x, 1/3)
\end{Sinput}
\begin{Soutput}
#> [caracas]: 1.11751906874186
\end{Soutput}
\end{Schunk}

In the first case, \(1/3\) is regarded as a fraction. In the second
case, \(1/3\) is evaluated numerically in R before \CRANpkg{caracas}
gets the value. As a consequence we have:

\begin{Schunk}
\begin{Sinput}
R> subs(f, x, "1/3 + 1/4")
\end{Sinput}
\begin{Soutput}
#> [caracas]:    / 49\
#>            exp|---|
#>               \144/
\end{Soutput}
\begin{Sinput}
R> subs(f, x, 1/3 + 1/4)
\end{Sinput}
\begin{Soutput}
#> [caracas]: 1.40533790799144
\end{Soutput}
\end{Schunk}

An exact evaluation can be evaluated numerically afterwards:

\begin{Schunk}
\begin{Sinput}
R> subs(f, x, "1/3 + 1/4") %>% as_expr()
\end{Sinput}
\begin{Soutput}
#> [1] 1.41
\end{Soutput}
\begin{Sinput}
R> subs(f, x, "1/3 + 1/4") %>% N(30) # Exact representation up to 30 decimals
\end{Sinput}
\begin{Soutput}
#> [caracas]: 1.40533790799143890537847414768
\end{Soutput}
\end{Schunk}

We can also convert the \CRANpkg{caracas} symbol to an R expression that
is subsequently evaluated:

\begin{Schunk}
\begin{Sinput}
R> f %>% as_expr()
\end{Sinput}
\begin{Soutput}
#> expression(exp(x^2))
\end{Soutput}
\begin{Sinput}
R> f %>% as_expr() %>% eval(list(x = 1/3 + 1/4))
\end{Sinput}
\begin{Soutput}
#> [1] 1.41
\end{Soutput}
\end{Schunk}

\hypertarget{taylor-expansion}{%
\subsubsection{Taylor expansion}\label{taylor-expansion}}

We perform a fourth order Taylor expansion of \(f(x) = \cos(x)\) around
\(x = 0\):

\begin{Schunk}
\begin{Sinput}
R> def_sym(x)
R> f <- cos(x)
R> ft_with_O <- taylor(f, x0 = 0, n = 4+1); ft_with_O
\end{Sinput}
\begin{Soutput}
#> [caracas]:      2    4          
#>                x    x     / 5.0\
#>            1 - -- + -- + O\x   /
#>                2    24
\end{Soutput}
\end{Schunk}

The order term can be removed:

\begin{Schunk}
\begin{Sinput}
R> ft <- drop_remainder(ft_with_O); ft
\end{Sinput}
\begin{Soutput}
#> [caracas]:  4    2    
#>            x    x     
#>            -- - -- + 1
#>            24   2
\end{Soutput}
\begin{Sinput}
R> ft %>% as_expr()
\end{Sinput}
\begin{Soutput}
#> expression(x^4/24 - x^2/2 + 1)
\end{Soutput}
\end{Schunk}

\hypertarget{matrix-algebra}{%
\subsubsection{Matrix algebra}\label{matrix-algebra}}

We briefly demonstrate the use matrices in \CRANpkg{caracas} (see also
\url{https://r-cas.github.io/caracas/}):

\begin{Schunk}
\begin{Sinput}
R> A <- matrix_(c("a", "b", "c", "d"), nrow = 2, ncol = 2) # Note the '_' postfix
R> # Or: matrix(c("a", "b", "c", "d"), nrow = 2, ncol = 2) %>% as_sym()
\end{Sinput}
\end{Schunk}

Note that \texttt{rbind()} and \texttt{cbind()} also works on
\CRANpkg{caracas} (vector) symbols:

\begin{Schunk}
\begin{Sinput}
R> c1 <- as_sym(c("a", "b"))
R> c2 <- as_sym(c("c", "d"))
R> A <- cbind(c1, c2); A
\end{Sinput}
\begin{Soutput}
#> [caracas]: [a  c]
#>            [    ]
#>            [b  d]
\end{Soutput}
\begin{Sinput}
R> D <- diag_(c("e1", "e2")); D # Note the '_' postfix
\end{Sinput}
\begin{Soutput}
#> [caracas]: [e1  0 ]
#>            [      ]
#>            [0   e2]
\end{Soutput}
\end{Schunk}

Some routines are demonstrated below:

\begin{Schunk}
\begin{Sinput}
R> detA <- det(A)
R> Ai <- inv(A) # Shorthand for solve_lin(A)
R> AD <- A %*% D
\end{Sinput}
\end{Schunk}

\[
\texttt{detA} = a d - b c, \quad
\texttt{Ai} = \left[\begin{matrix}\frac{d}{a d - b c} & - \frac{c}{a d - b c}\\- \frac{b}{a d - b c} & \frac{a}{a d - b c}\end{matrix}\right], \quad
\texttt{AD} = \left[\begin{matrix}a e_{1} & c e_{2}\\b e_{1} & d e_{2}\end{matrix}\right]
\]

\begin{Schunk}
\begin{Sinput}
R> evec <- eigenvec(A)
R> evec1 <-evec[[1]]$eigvec %>% simplify()
R> eval <- eigenval(A)
R> eval1 <- eval[[1]]$eigval %>% simplify()
\end{Sinput}
\end{Schunk}

\[
\texttt{evec1} = \left[\begin{matrix}- \frac{2 c}{a - d + \sqrt{a^{2} - 2 a d + 4 b c + d^{2}}}\\1\end{matrix}\right], \quad
\texttt{eval1} = \frac{a}{2} + \frac{d}{2} - \frac{\sqrt{a^{2} - 2 a d + 4 b c + d^{2}}}{2}.
\]

\begin{Schunk}
\begin{Sinput}
R> B <- matrix_(c("b", "0", "0", "1"), nrow = 2, ncol = 2)
R> qr_res <- QRdecomposition(B)
R> Q <- qr_res$Q 
R> R <- qr_res$R
\end{Sinput}
\end{Schunk}

\[
\texttt{Q} = \left[\begin{matrix}\frac{b}{\left|{b}\right|} & 0\\0 & 1\end{matrix}\right], \quad
\texttt{R} = \left[\begin{matrix}\left|{b}\right| & 0\\0 & 1\end{matrix}\right].
\]

\hypertarget{lagrange-multiplier-and-maximizing-a-likelihood}{%
\subsubsection{Lagrange multiplier and maximizing a
likelihood}\label{lagrange-multiplier-and-maximizing-a-likelihood}}

Here we illustrate how to maximize a multinomial likelihood using
Lagrange multiplier. Consider a multinomial model with three categories
with probabilities \(p_1\), \(p_2\) and \(p_3\) such that
\(p_1 + p_2 + p_3 = 1\). We then observe counts \(y_1\), \(y_2\) and
\(y_3\) of each category. The multinomial log-likelihood for this model
is \begin{align}
  l(p) &= y_1 \log(p_1) + y_2 \log(p_2) + y_3 \log(p_3) .
\end{align}

We wish to maximize \(l(p)\) under the constraint that
\(p_1 + p_2 + p_3 = 1\). This can be achieved using Lagrange multiplier
where we instead solve the unconstrained optimization problem
\(\max_p L(p)\) where \begin{align}
  L(p) &= -l(p) + \lambda g(p) \quad \text{under the constraint that} \\
  g(p) &= p_1 + p_2 + p_3 - 1 = 0.
\end{align}

The function \(L\) can be expressed in \CRANpkg{caracas} as follows
where we create character vectors in R and convert them to a
\CRANpkg{caracas} symbol using \texttt{as\_sym()}:

\begin{Schunk}
\begin{Sinput}
R> p <- as_sym(paste0("p", 1:3))
R> y <- as_sym(paste0("y", 1:3))
R> def_sym(a) 
R> l <- sum(y * log(p))
R> L <- -l + a * (sum(p) - 1); L
\end{Sinput}
\begin{Soutput}
#> [caracas]: a*(p1 + p2 + p3 - 1) - y1*log(p1) - y2*log(p2) - y3*log(p3)
\end{Soutput}
\end{Schunk}

To solve the unconstrained optimization problem we find the critical
points and afterwards check the eigenvalues of the Hessian (at the
critical points). The critical points are found as follows by first
finding the gradient with \texttt{der()} (for derivative) and then
equating the gradient to zero:

\begin{Schunk}
\begin{Sinput}
R> gL <- der(L, list(p, a))
\end{Sinput}
\end{Schunk}

Hence, \CRANpkg{caracas} computes the gradient to be \[
  \nabla L (p_1, p_2, p_3, a) = \left[\begin{matrix}a - \frac{y_{1}}{p_{1}} & a - \frac{y_{2}}{p_{2}} & a - \frac{y_{3}}{p_{3}} & p_{1} + p_{2} + p_{3} - 1\end{matrix}\right] .
\]

Next we solve \(\nabla L (p_1, p_2, p_3, a) = 0\):

\begin{Schunk}
\begin{Sinput}
R> sols <- solve_sys(gL, list(p, a)) # takes an RHS argument which defaults to zero
R> sols
\end{Sinput}
\begin{Soutput}
#> Solution 1:
#>   p1 =       y1     
#>        ------------
#>        y1 + y2 + y3 
#>   p2 =       y2     
#>        ------------
#>        y1 + y2 + y3 
#>   p3 =       y3     
#>        ------------
#>        y1 + y2 + y3 
#>   a  =  y1 + y2 + y3
\end{Soutput}
\end{Schunk}

One critical point is found. (Notice that in general it is difficult to
know how many critical points a function has. We will not go into
details about this aspect of the problem.) Next we verify that we have
found a minimum: We find the Hessian as a symbol and evaluate it in the
critical point:

\begin{Schunk}
\begin{Sinput}
R> H <- der2(l, p) # der2(...) is shorthand for calling der() twice
R> H_sol <- subs_lst(H, sols[[1]]) # Substitute solution into H
\end{Sinput}
\end{Schunk}

\[
  H = \left[\begin{matrix}- \frac{y_{1}}{p_{1}^{2}} & 0 & 0\\0 & - \frac{y_{2}}{p_{2}^{2}} & 0\\0 & 0 & - \frac{y_{3}}{p_{3}^{2}}\end{matrix}\right], \quad
  H_{\text{sol}} = \left[\begin{matrix}- \frac{\left(y_{1} + y_{2} + y_{3}\right)^{2}}{y_{1}} & 0 & 0\\0 & - \frac{\left(y_{1} + y_{2} + y_{3}\right)^{2}}{y_{2}} & 0\\0 & 0 & - \frac{\left(y_{1} + y_{2} + y_{3}\right)^{2}}{y_{3}}\end{matrix}\right]
\]

We verify that the solution is indeed a minimum: As \(H_{\text{sol}}\)
is a diagonal matrix, its eigenvalues are the diagonal entries. (In
general, eigenvalues/eigenvectors can be found using
\texttt{eigenval()}/\texttt{eigenvec()}.) Provided that all \(y_i > 0\),
all eigenvalues are negative and the likelihood reached a maximum.

\hypertarget{extending-calling-sympy-functions-directly}{%
\subsubsection{\texorpdfstring{Extending \CRANpkg{caracas} -- calling
SymPy functions
directly}{Extending  -- calling SymPy functions directly}}\label{extending-calling-sympy-functions-directly}}

The \CRANpkg{caracas} can be extended as it is possible to call SymPy
functions directly with the \texttt{sympy\_func()} function. Please
refer to the SymPy documentation at
\url{https://docs.sympy.org/latest/index.html} for documentation of
SymPy functions.

At the time of writing, the SymPy functions \texttt{collect()} and
\texttt{factor()} are not implemented in \CRANpkg{caracas}, but they can
be invoked as shown in the following. For example, we can collect terms
in a polynomial expression:

\begin{Schunk}
\begin{Sinput}
R> def_sym(x, y, z)
R> p <- x*y + x - 3 + 2*x^2 - z*x^2 + x^3
R> p %>% sympy_func("collect", x)
\end{Sinput}
\begin{Soutput}
#> [caracas]:  3    2                        
#>            x  + x *(2 - z) + x*(y + 1) - 3
\end{Soutput}
\end{Schunk}

We can also expand and factor a polynomial:

\begin{Schunk}
\begin{Sinput}
R> def_sym(x)
R> p <- (x - 1)^7
R> q <- p %>% expand(); q # or p %>% sympy_func("expand")
\end{Sinput}
\begin{Soutput}
#> [caracas]:  7      6       5       4       3       2          
#>            x  - 7*x  + 21*x  - 35*x  + 35*x  - 21*x  + 7*x - 1
\end{Soutput}
\begin{Sinput}
R> q %>% sympy_func("factor")
\end{Sinput}
\begin{Soutput}
#> [caracas]:        7
#>            (x - 1)
\end{Soutput}
\end{Schunk}

In passing we illustrate a difference between symbolic and numerical
mathematics: Floating point arithmetic can lead to catastrophic
cancelations when nearly identical quantities are subtracted. This can
be demonstrated as follows. Evaluate \texttt{p} and \texttt{q} on a
range of \texttt{x}-values near 1, and plot the results in Fig.
\ref{fig:pols-pq}.

\begin{Schunk}
\begin{figure}
\includegraphics{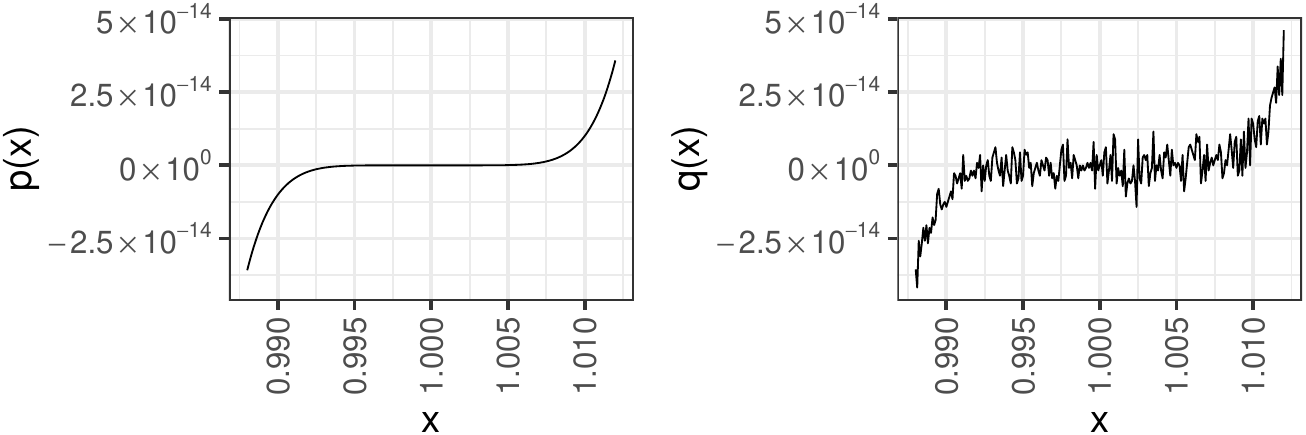} \caption[Difference between symbolic and numerical mathematics]{Difference between symbolic and numerical mathematics. Left: Plot of polynomial $p(x) = (x-1)^7$. Right: Plot of explansion of same polynomial.}\label{fig:pols-pq}
\end{figure}
\end{Schunk}

\hypertarget{statistics-examples}{%
\subsection{Statistics examples}\label{statistics-examples}}

In a linear model setting where \(\mathbf{E}(y)=X\beta\), the least
squares estimate of \(\beta\) can be written as
\(\hat\beta = (X^\top X)^{-1} X^\top y\). It is illustrative to use
symbolic mathematics for illustrating what is computed in the matrix
algebra.

\hypertarget{one-way-analysis-of-variance-one-way-anova}{%
\subsubsection{One-way analysis of variance (one-way
ANOVA)}\label{one-way-analysis-of-variance-one-way-anova}}

First consider one-way analysis of variance (ANOVA).

\begin{Schunk}
\begin{Sinput}
R> ngrp <- 3 # Number of groups
R> spg  <- 2 # Number of subjects per group
R> g <- seq_len(ngrp)
R> f <- factor(rep(g, each = spg))
R> y <- as_sym(paste0("y", seq_along(f)))
R> X <- as_sym(model.matrix(~ f))
\end{Sinput}
\end{Schunk}

We compute the usual quantities needed for finding the least squares
estimate for the regression coefficients.

\begin{Schunk}
\begin{Sinput}
R> XtX <- t(X) %*% X
R> XtXinv <- inv(XtX) # Shorthand for solve_lin(XtX)
\end{Sinput}
\end{Schunk}

\[
  X = \left[\begin{matrix}1 & 0 & 0\\1 & 0 & 0\\1 & 1 & 0\\1 & 1 & 0\\1 & 0 & 1\\1 & 0 & 1\end{matrix}\right], \quad
  X^\top X = \left[\begin{matrix}6 & 2 & 2\\2 & 2 & 0\\2 & 0 & 2\end{matrix}\right], \quad
  (X^\top X)^{-1} = \left[\begin{matrix}\frac{1}{2} & - \frac{1}{2} & - \frac{1}{2}\\- \frac{1}{2} & 1 & \frac{1}{2}\\- \frac{1}{2} & \frac{1}{2} & 1\end{matrix}\right] 
\]

Likewise,

\begin{Schunk}
\begin{Sinput}
R> Xty <- t(X) %*% y
R> beta_hat <- XtXinv %*% Xty
R> y_hat <- X %*% beta_hat
\end{Sinput}
\end{Schunk}

\[
  X^\top y = \left[\begin{matrix}y_{1} + y_{2} + y_{3} + y_{4} + y_{5} + y_{6}\\y_{3} + y_{4}\\y_{5} + y_{6}\end{matrix}\right], \quad
  \hat{\beta} = \frac 1 2 \left[\begin{matrix}y_{1} + y_{2}\\- y_{1} - y_{2} + y_{3} + y_{4}\\- y_{1} - y_{2} + y_{5} + y_{6}\end{matrix}\right], \quad 
  \hat{y} = \frac 1 2 \left[\begin{matrix}y_{1} + y_{2}\\y_{1} + y_{2}\\y_{3} + y_{4}\\y_{3} + y_{4}\\y_{5} + y_{6}\\y_{5} + y_{6}\end{matrix}\right]
\]

Hence \(X^\top y\) consists of the sum of all observations, the sum of
observations in group 2 and the sum of observations in group 3.
Similarly, \(\hat\beta\) consists of the average in group 1, the average
in group 2 minus the average in group 1 and the average in group 3 minus
the average in group 1. Fitted values are simply group averages.

Next consider a linear model setting,
i.e.~\(y\sim N(X\beta, \sigma^2 I)\). The score function and Hessian
matrix can be derived in closed form as follows: Define residuals
\(r=y - X \hat{\beta}\) and residual-sum-of-squares as
\(RSS=\sum_i r_i^2\). The log-likelihood based on \(n\) observations is
\[
 l = - \frac n 2 \sigma^2 - \frac 1 {2\sigma^2} RSS.
\]

Here we can find critical points of \(l\) for \(RSS\) and \(\sigma^2\)
independently, so we ignore \(\sigma^2\) and proceed focusing on \(RSS\)
as follows:

\begin{Schunk}
\begin{Sinput}
R> beta <- as_sym(paste0("beta", 1:3))
R> res <- y - X %*% beta
R> RSS <- sum(res^2) 
R> logL <- - RSS / 2
\end{Sinput}
\end{Schunk}

We find the score function and Hessian matrix by differentiation.

\begin{Schunk}
\begin{Sinput}
R> Score <- der(logL, beta) %>% matrify()    # Convert Python list to vector
R> Hessian <- der2(logL, beta) %>% matrify() # Convert Python list to matrix
\end{Sinput}
\end{Schunk}

\[
\texttt{Score}= \left[\begin{matrix}- 6 \beta_{1} - 2 \beta_{2} - 2 \beta_{3} + y_{1} + y_{2} + y_{3} + y_{4} + y_{5} + y_{6}\\- 2 \beta_{1} - 2 \beta_{2} + y_{3} + y_{4}\\- 2 \beta_{1} - 2 \beta_{3} + y_{5} + y_{6}\end{matrix}\right], \quad 
\texttt{Hessian}= \left[\begin{matrix}-6 & -2 & -2\\-2 & -2 & 0\\-2 & 0 & -2\end{matrix}\right]
\]

Notice the following: The output from \texttt{der} and \texttt{der2} are
lists in Python, and to be able to work with these quantities as we
normally do a coercion to matrices is needed. The \texttt{matrify()}
function does this. We conclude this example by solving the likelihood
equations which, fortunately, leads to the same quantity as
\(\hat\beta\) derived above:

\begin{Schunk}
\begin{Sinput}
R> sol <- solve_sys(Score, beta)
R> sol
\end{Sinput}
\begin{Soutput}
#> Solution 1:
#>   beta1 =  y1   y2
#>           -- + --
#>           2    2 
#>   beta2 =    y1   y2   y3   y4
#>           - -- - -- + -- + --
#>             2    2    2    2 
#>   beta3 =    y1   y2   y5   y6
#>           - -- - -- + -- + --
#>             2    2    2    2
\end{Soutput}
\end{Schunk}

Similar considerations can be done for other linear models, e.g.~ a
(balanced) two-way analysis of variance (two-way ANOVA).

\hypertarget{probabilistic-principal-component-analysis}{%
\subsubsection{Probabilistic principal component
analysis}\label{probabilistic-principal-component-analysis}}

A probabilistic principal component analysis (PCA) model arises as
follows, see e.g. \citet{bishop:06}, pp.~570: There is a latent vector
\(z\) and it is assumed that \(z\sim N(0,I)\). There is a vector \(x\)
of observables and it is assumed that
\(x \mid z \sim N(Wz + \mu, v^2I)\). It is not a restriction to assume
that \(\mu=0\) because we center each variable around its average. We
can write the model as \begin{align} \label{eq:ppca-model}
z = e_z \quad \text{and} \quad
x = Wz + e_x, 
\end{align} where \(e_z \sim N(0,I)\) and \(e_x\sim N(0,v^2I)\) are
error terms that are assumed independent. The \(W\) matrix is the model
matrix of weights that reflects model assumptions about how \(z\)
impacts \(x\).

In a statistical inference setting, the unknown parameters (\(v^2\) and
the components of \(W\)) must be estimated. To do this the covariance
matrix and concentration matrix are needed. The first step is to
identify the structural form of the covariance matrix
\(V=\mathbf{Var}(z,x)\) and concentration matrix \(K=V^{-1}\), and
\CRANpkg{caracas} can do this for us. The next step is to use these
structural forms in the likelihood function which can then be maximized.

Define \[
V   = \mathbf{Var}(z, x)= \left(
    \begin{array}{cccccccccccccccccccccccccccccccccccc}
      V_{zz} & V_{zx} \\ V_{xz} & V_{xx} \\
    \end{array}
  \right), \quad
K   = V^{-1}= \left(
    \begin{array}{cccccccccccccccccccccccccccccccccccc}
      K_{zz} & K_{zx} \\ K_{xz} & K_{xx} \\
    \end{array}
  \right), \quad
V_e = \mathbf{Var}(e)   = \left(
    \begin{array}{cccccccccccccccccccccccccccccccccccc}
      I & 0 \\ 0 & v^2 I \\
    \end{array}
  \right) .
\] Also define \begin{align} \label{eq:ppca-L}
L = \left(
    \begin{array}{cccccccccccccccccccccccccccccccccccc}
      I & 0 \\ -W & I \\
    \end{array}
  \right) \mbox{ and note that }
  L^{-1} = \left(
    \begin{array}{cccccccccccccccccccccccccccccccccccc}
      I & 0 \\ W & I \\
    \end{array}
  \right).
\end{align} Isolating error terms in \eqref{eq:ppca-model} gives \[
e 
= \left(
    \begin{array}{cccccccccccccccccccccccccccccccccccc}
      e_z \\ e_x \\
    \end{array}
  \right) 
= \left(
    \begin{array}{cccccccccccccccccccccccccccccccccccc}
      I & 0 \\ -W & I \\
    \end{array}
  \right) \left(
    \begin{array}{cccccccccccccccccccccccccccccccccccc}
      z \\ x \\
    \end{array}
  \right) 
= L \left(
    \begin{array}{cccccccccccccccccccccccccccccccccccc}
      z \\ x \\
    \end{array}
  \right) .
\] Hence \(\mathbf{Var}(e) = L \mathbf{Var}(z,x) L^\top\) and therefore

\begin{equation}
V = \mathbf{Var}(z,x)=L^{-1}\mathbf{Var}(e) (L^{-1})^\top\mbox{ and }
K = V^{-1}= L^\top\mathbf{Var}(e)^{-1}L.
\label{eq:KV}
\end{equation}

Since the error terms are independent, a direct calculation gives
\begin{align}
  \label{eq:ppca1}
  V &= \left(
    \begin{array}{cccccccccccccccccccccccccccccccccccc}
      I & W^\top\\ W & WW^\top+ v^2I \\
    \end{array}
  \right) \mbox{ and }
  K = \left(
    \begin{array}{cccccccccccccccccccccccccccccccccccc}
      I + v^{-2}W^\top W & -v^{-2}W^\top\\ v^{-2}W & v^{-2}I  \\
    \end{array}
  \right)
\end{align}

The following observations can be made:

\begin{enumerate}
\def\labelenumi{\arabic{enumi}.}
\item
  First recall a general result on the multivariate normal distribution.
  Suppose \(U=(U_1, \dots, U_d)\sim N(\mu, V)\), and let \(K=V^{-1}\).
  Then \(K_{ij} = 0\) if and only if \(U_i\) and \(U_j\) are
  conditionally independent given all other components of \(U\). Next
  return to the specific setting. The lower right corner, \(K_{xx}\), of
  \(K\) is \(v^{-2}I\) and the fact that this matrix is diagonal
  reflects that all pairs of observables \(x_u\) and \(x_v\) are
  conditionally independent given the latent variables \(z\).
\item
  The lower right corner, \(V_{xx}\), of \(V\) is \(WW^\top+ v^2I\) and
  this matrix is the covariance matrix of observables \(x\). The inverse
  of \(WW^\top+ v^2I\) is the concentration matrix of \(x\) (in the
  marginal distribution of \(x\)) and this concentration matrix does not
  in general contain zeros. There are no conditional independencies
  among the observables alone; conditional independencies arise from
  conditioning on the latent variables.
\item
  To estimate the parameters \(W\) and \(v^2\) we can maximize the
  log--likelihood for the observables with covariance matrix
  \(V_{obs}=WW^\top+ v^2I\). This can often be done directly using
  \texttt{optim()}. (Notice that we can just center data to eliminate
  the parameter \(\mu\)).
\end{enumerate}

\begin{Schunk}
\begin{figure}

{\centering \includegraphics{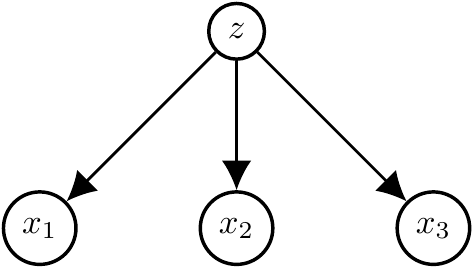} 

}

\caption[A directed acyclic graph (DAG) illustrating probabilistic PCA]{A directed acyclic graph (DAG) illustrating probabilistic PCA:  The observables $x=(x_1, x_2, x_3)$ are conditionally independent given $z$.}\label{fig:ppca1}
\end{figure}
\end{Schunk}

\hypertarget{a-simple-example}{%
\paragraph{A simple example}\label{a-simple-example}}

A particularly simple example is the following where \(z\) is
one--dimensional and \(x\) is three--dimensional and \(W\) is a
\(3\times 1\) matrix with \(a\) in all entries: \[
x_i = a z + e_i, \quad i=1, \dots, 3, \quad z = u
\] All \(e_1, \dots, e_3\) are \(N(0,v^2)\) distributed,
\(u \sim N(0, 1)\) and all error terms are independent. See an
illustration of this model in Fig. \ref{fig:ppca1}. Let
\(e=(e, \dots, e_3)\) and \(x=(x_1, \dots x_3)\). Hence
\(e \sim N(0, v^2 I)\) and \(\mathbf{Var}(u,e)\) is a diagonal matrix,
\(V_{ue}=\mathbf{diag}(1, v^2, \dots, v^2)\). Isolating error terms
gives \[
(u,e)= \left[\begin{matrix}u\\e_{1}\\e_{2}\\e_{3}\end{matrix}\right] = \left[\begin{matrix}1 & 0 & 0 & 0\\- a & 1 & 0 & 0\\- a & 0 & 1 & 0\\- a & 0 & 0 & 1\end{matrix}\right] \left[\begin{matrix}z\\x_{1}\\x_{2}\\x_{3}\end{matrix}\right] = L (z,x), \mbox{ say. }
\]

\begin{Schunk}
\begin{Sinput}
R> N <- 3
R> L <- diag_("1", N + 1)
R> L[cbind(1 + (1:N), 1)] <- "-a"
R> Vue <- matrix_("0", nrow = N + 1, ncol = N + 1)
R> diag(Vue) <- c("1", rep("v2", N))
\end{Sinput}
\end{Schunk}

Following (\ref{eq:KV}) we find \(V\) and \(K\) as:

\begin{Schunk}
\begin{Sinput}
R> V <- inv(L) %*% Vue %*% t(inv(L))
R> K <- t(L) %*% inv(Vue) %*% L
\end{Sinput}
\end{Schunk}

\[
V = \left[\begin{matrix}1 & a & a & a\\a & a^{2} + v_{2} & a^{2} & a^{2}\\a & a^{2} & a^{2} + v_{2} & a^{2}\\a & a^{2} & a^{2} & a^{2} + v_{2}\end{matrix}\right], \quad K = \left[\begin{matrix}\frac{3 a^{2}}{v_{2}} + 1 & - \frac{a}{v_{2}} & - \frac{a}{v_{2}} & - \frac{a}{v_{2}}\\- \frac{a}{v_{2}} & \frac{1}{v_{2}} & 0 & 0\\- \frac{a}{v_{2}} & 0 & \frac{1}{v_{2}} & 0\\- \frac{a}{v_{2}} & 0 & 0 & \frac{1}{v_{2}}\end{matrix}\right]
\]

\hypertarget{introducing-data}{%
\paragraph{Introducing data}\label{introducing-data}}

Let \(S\) is the empirical covariance matrix for the observed variables
based on \(n\) observations. The observed-data log--likelihood is \[
  l 
  = \frac{n}{2} (\log \mathbf{det}(V_{xx}^{-1}) - \mathbf{tr}(V_{xx}^{-1} S))
  = \frac{n}{2} (\log \mathbf{det}(K_{xx}) - \mathbf{tr}(K_{xx} S)).
\]

Now \(K_{xx}= V_{xx}^{-1}\) can be extracted and used in the likelihood
function:

\begin{Schunk}
\begin{Sinput}
R> Vxx_inv <- inv(V[-1, -1])
\end{Sinput}
\end{Schunk}

Alternatively, \(K_{xx}\) can be found as

\begin{Schunk}
\begin{Sinput}
R> Kxx <- (K[-1, -1] - K[-1, 1, drop=F] %*% 
+           inv(K[1, 1, drop=F]) %*% K[1, -1, drop=F]) %>% simplify
\end{Sinput}
\end{Schunk}

\[
 K_{xx} = \frac{1}{v_{2} \left(3 a^{2} + v_{2}\right)} \left[\begin{matrix}2 a^{2} + v_{2} & - a^{2} & - a^{2}\\- a^{2} & 2 a^{2} + v_{2} & - a^{2}\\- a^{2} & - a^{2} & 2 a^{2} + v_{2}\end{matrix}\right]
\]

It remains to be investigated in practice whether it is computationally
more efficient to construct \(V_{xx}\) numerically first and then invert
\(V_{xx}\) to obtain \(K_{xx}\) numerically instead of finding
\(K_{xx}\) symbolically and subsequently evaluating this numerically.

\hypertarget{with-a-view-towards-teaching}{%
\subsection{With a view towards
teaching}\label{with-a-view-towards-teaching}}

We have found ANOVA examples (and variants hereof) useful in connection
with teaching: Students are often exposed to estimating the vector of
regression coefficients as \[
\hat\beta = (X^\top X)^{-1} X^\top y.
\]

However, in the computer area, students are not always exposed to what
is really computed in simple cases: Various group sums and differences
of these. It is often illustrative for students to study, for example,
the effect of: (1) including/excluding the intercept term in a model,
(2) working with different contrasts in the models, and (3) making a
design imbalanced.

For the probabilistic PCA example it is illustrative for students (1) to
take the step from the symbolic model formulation to estimation (using
e.g.~\texttt{optim()}). (2) It can also be illustrative to realize the
interpretation implied by forcing certain elements of \(W\) to be
identical (as the case was above). (3) Likewise, it is straight forward
to change the conditional variance of \(x\) given \(z\) from
\(\sigma^2I\) to a diagonal matrix \(\Psi\) (which is the case in
standard factor analysis). Elaborting further, it is illustrative for
students to see for example and autoregression and a dynamic linear
model formulated in similar way.

\hypertarget{discussion}{%
\subsection{Discussion}\label{discussion}}

We have presented the \CRANpkg{caracas} package and argued that the
package extends the functionality of R significantly with respect to
symbolic mathematics. One practical virtue of \CRANpkg{caracas} is that
the package integrates nicely with \CRANpkg{Rmarkdown},
\citet{rmarkdown}, (e.g.~with the \texttt{tex()} functionality) and thus
supports creating of scientific documents and teaching material. As for
the usability in practice we await feedback from users.

\hypertarget{acknowledgements}{%
\subsection{Acknowledgements}\label{acknowledgements}}

We would like to thank the R foundation for financial support for
creating the \CRANpkg{caracas} package, users for pin pointing points
that can be improved in \CRANpkg{caracas} and Ege Rubak (Aalborg
University, Denmark) and Malte Bødkergaard Nielsen (Aalborg University,
Denmark) for comments on this manuscript.

\bibliography{RJreferences}

\begin{thebibliography}{11}
\providecommand{\natexlab}[1]{#1}
\providecommand{\url}[1]{\texttt{#1}}
\expandafter\ifx\csname urlstyle\endcsname\relax
  \providecommand{\doi}[1]{doi: #1}\else
  \providecommand{\doi}{doi: \begingroup \urlstyle{rm}\Url}\fi

\bibitem[Allaire et~al.(2021)Allaire, Xie, McPherson, Luraschi, Ushey, Atkins,
  Wickham, Cheng, Chang, and Iannone]{rmarkdown}
J.~Allaire, Y.~Xie, J.~McPherson, J.~Luraschi, K.~Ushey, A.~Atkins, H.~Wickham,
  J.~Cheng, W.~Chang, and R.~Iannone.
\newblock \emph{rmarkdown: Dynamic Documents for R}, 2021.
\newblock URL \url{https://github.com/rstudio/rmarkdown}.
\newblock R package version 2.7.

\bibitem[Andersen and Højsgaard(2019)]{ryacas}
M.~M. Andersen and S.~Højsgaard.
\newblock {Ryacas: A computer algebra system in R}.
\newblock \emph{Journal of Open Source Software}, 4\penalty0 (42), 2019.
\newblock URL \url{https://doi.org/10.21105/joss.01763}.

\bibitem[Bache and Wickham(2020)]{magrittr}
S.~M. Bache and H.~Wickham.
\newblock \emph{{magrittr: A Forward-Pipe Operator for R}}, 2020.
\newblock URL \url{https://CRAN.R-project.org/package=magrittr}.
\newblock R package version 2.0.1.

\bibitem[Bishop(2006)]{bishop:06}
C.~M. Bishop.
\newblock \emph{Pattern Recognition and Machine Learning}.
\newblock Springer, New York, USA, 2006.

\bibitem[Kaplan et~al.(2020)Kaplan, Pruim, and Horton]{mosaicCalc}
D.~T. Kaplan, R.~Pruim, and N.~J. Horton.
\newblock \emph{{mosaicCalc: Function-Based Numerical and Symbolic
  Differentiation and Antidifferentiation}}, 2020.
\newblock URL \url{https://CRAN.R-project.org/package=mosaicCalc}.
\newblock R package version 0.5.1.

\bibitem[Meurer et~al.(2017)Meurer, Smith, Paprocki, \v{C}ert\'{i}k, Kirpichev,
  Rocklin, Kumar, Ivanov, Moore, Singh, Rathnayake, Vig, Granger, Muller,
  Bonazzi, Gupta, Vats, Johansson, Pedregosa, Curry, Terrel, Rou\v{c}ka, Saboo,
  Fernando, Kulal, Cimrman, and Scopatz]{sympy}
A.~Meurer, C.~P. Smith, M.~Paprocki, O.~\v{C}ert\'{i}k, S.~B. Kirpichev,
  M.~Rocklin, A.~Kumar, S.~Ivanov, J.~K. Moore, S.~Singh, T.~Rathnayake,
  S.~Vig, B.~E. Granger, R.~P. Muller, F.~Bonazzi, H.~Gupta, S.~Vats,
  F.~Johansson, F.~Pedregosa, M.~J. Curry, A.~R. Terrel, v.~Rou\v{c}ka,
  A.~Saboo, I.~Fernando, S.~Kulal, R.~Cimrman, and A.~Scopatz.
\newblock Sympy: symbolic computing in python.
\newblock \emph{PeerJ Computer Science}, 3:\penalty0 e103, Jan. 2017.
\newblock ISSN 2376-5992.
\newblock \doi{10.7717/peerj-cs.103}.
\newblock URL \url{https://doi.org/10.7717/peerj-cs.103}.

\bibitem[Pinkus et~al.(2016)Pinkus, Winnitzky, and Mazur]{yacas}
A.~Pinkus, S.~Winnitzky, and G.~Mazur.
\newblock Yacas - yet another computer algebra system.
\newblock Technical report, 2016.
\newblock URL \url{https://yacas.readthedocs.io/en/latest/}.

\bibitem[Pinkus and Winitzki(2002)]{Pinkus2002}
A.~Z. Pinkus and S.~Winitzki.
\newblock {YACAS: A Do-It-Yourself Symbolic Algebra Environment}.
\newblock In \emph{Proceedings of the Joint International Conferences on
  Artificial Intelligence, Automated Reasoning, and Symbolic Computation}, AISC
  '02/Calculemus '02, pages 332--336, London, UK, UK, 2002. Springer-Verlag.
\newblock ISBN 3-540-43865-3.
\newblock \doi{10.1007/3-540-45470-5_29}.
\newblock URL \url{http://doi.org/10.1007/3-540-45470-5_29}.

\bibitem[{R Core Team}(2021)]{R}
{R Core Team}.
\newblock \emph{R: A Language and Environment for Statistical Computing}.
\newblock R Foundation for Statistical Computing, Vienna, Austria, 2021.
\newblock URL \url{http://www.R-project.org/}.
\newblock {ISBN} 3-900051-07-0.

\bibitem[{United Nations General Assembly}(2015)]{UN17}
{United Nations General Assembly}.
\newblock Sustainable development goals, 2015.
\newblock \url{https://sdgs.un.org/}.

\bibitem[Ushey et~al.(2020)Ushey, Allaire, and Tang]{reticulate}
K.~Ushey, J.~Allaire, and Y.~Tang.
\newblock \emph{reticulate: Interface to 'Python'}, 2020.
\newblock URL \url{https://CRAN.R-project.org/package=reticulate}.
\newblock R package version 1.18.

\end{thebibliography}

\hypertarget{appendix-technicalities}{%
\subsection{Appendix: Technicalities}\label{appendix-technicalities}}

To avoid confusion we elaborate on the construction of symbols in the
following:

\begin{itemize}
\tightlist
\item
  \texttt{as\_sym()} converts an R object (or string) to symbol.
\item
  \texttt{symbol()} declares a symbol by string, and allows for
  assumptions.
\item
  \texttt{def\_sym()} declares a symbol (either by string or
  non-standard evaluation) and assigns to an R variable with same name.
\end{itemize}

The behaviour of \texttt{def\_sym()} can be obtained by both
\texttt{symbol()} and \texttt{as\_sym()}, but the two latter require an
explicit assignment. Thus the following three statements are equivalent:

\begin{Schunk}
\begin{Sinput}
R> a <- as_sym("a")
R> a <- symbol("a")
R> def_sym(a)
\end{Sinput}
\end{Schunk}

To elaborate, consider a vector in R: Using \texttt{symbol} - the
following fails because \texttt{a} is an \texttt{a} object and not a
string:

\begin{Schunk}
\begin{Sinput}
R> a <- c(-1, 1); a
\end{Sinput}
\begin{Soutput}
#> [1] -1  1
\end{Soutput}
\begin{Sinput}
R> a <- symbol(a)  
\end{Sinput}
\begin{Soutput}
#> Error in verify_variable_name(x): The name must have length 1
\end{Soutput}
\end{Schunk}

On the other hand, \texttt{as\_sym} works as expected. Using
\texttt{def\_sym} also works, but not as the user expects: A new
variable \texttt{a} is created and the old \texttt{a} (the vector) is no
longer bound to a variable:

\begin{Schunk}
\begin{Sinput}
R> a2 <- as_sym(a); t(a2)
\end{Sinput}
\begin{Soutput}
#> [caracas]: [-1  1]
\end{Soutput}
\begin{Sinput}
R> a  ## a is unchanged
\end{Sinput}
\begin{Soutput}
#> [1] -1  1
\end{Soutput}
\begin{Sinput}
R> def_sym(a); a
\end{Sinput}
\begin{Soutput}
#> [caracas]: a
\end{Soutput}
\end{Schunk}

\hypertarget{appendix-assumptions}{%
\subsubsection{Appendix: Assumptions}\label{appendix-assumptions}}

It is sometimes required to impose assumptions on variables. There is
currently (limited) support for this in \CRANpkg{caracas}:

\begin{Schunk}
\begin{Sinput}
R> x <- symbol("x")
R> sol <- solve_sys(x^2 + 1, x); sol
\end{Sinput}
\begin{Soutput}
#> Solution 1:
#>   x =  -I 
#> Solution 2:
#>   x =  I
\end{Soutput}
\end{Schunk}

Requiring \texttt{x} to be real:

\begin{Schunk}
\begin{Sinput}
R> x <- symbol("x", real = TRUE)
R> ask(x, 'real')
\end{Sinput}
\begin{Soutput}
#> [1] TRUE
\end{Soutput}
\begin{Sinput}
R> sol <- solve_sys(x^2 + 1, x); sol
\end{Sinput}
\begin{Soutput}
#> No solutions
\end{Soutput}
\end{Schunk}

Requiring \texttt{x} to be positive:

\begin{Schunk}
\begin{Sinput}
R> x <- symbol("x", positive = TRUE)
R> ask(x, 'positive')
\end{Sinput}
\begin{Soutput}
#> [1] TRUE
\end{Soutput}
\begin{Sinput}
R> sol <- solve_sys(x^2 - 1, x); sol
\end{Sinput}
\begin{Soutput}
#> Solution 1:
#>   x =  1
\end{Soutput}
\end{Schunk}

\address{%
Mikkel Meyer Andersen\\
Department of Mathematical Sciences, Aalborg University, Denmark\\%
Skjernvej 4A\\ 9220 Aalborg Ø, Denmark\\
\\\href{mailto:mikl@math.aau.dk}{\nolinkurl{mikl@math.aau.dk}}
}

\address{%
Søren Højsgaard\\
Department of Mathematical Sciences, Aalborg University, Denmark\\%
Skjernvej 4A\\ 9220 Aalborg Ø, Denmark\\
\\\href{mailto:sorenh@math.aau.dk}{\nolinkurl{sorenh@math.aau.dk}}
}

\end{document}